\documentclass[12pt,preprint]{aastex}
\usepackage{natbib}

\begin{document}

\shortauthors{Gordon et al.}
\shorttitle{MIRI Operations \& Data Reduction}

\title{The Mid-Infrared Instrument for the James Webb Space Telescope, X. Operations and Data Reduction}

\author{Karl~D.~Gordon\altaffilmark{1,2}, 
   C. H. ~Chen\altaffilmark{1},
   Rachel E.~Anderson\altaffilmark{1},
   Ruym\'an~Azzollini\altaffilmark{3,4},
   L.~Bergeron\altaffilmark{1},
   Patrice~Bouchet\altaffilmark{5},   
   Jeroen~Bouwman\altaffilmark{6},
   Misty~Cracraft\altaffilmark{1},
   Sebastian~Fischer\altaffilmark{7,8},
   Scott~D.~Friedman\altaffilmark{1},
   Macarena Garc\'ia-Mar\'in\altaffilmark{8},
   Alistair~Glasse\altaffilmark{9},
   Adrian~M.~Glauser\altaffilmark{10},
   G. B. Goodson\altaffilmark{11},
   T. P. Greene\altaffilmark{12},
   Dean~C.~Hines\altaffilmark{1},
   M. A. Khorrami\altaffilmark{11},
   Fred~Lahuis\altaffilmark{13,14},
   C.-P.~Lajoie\altaffilmark{1},
   M. E.~Meixner\altaffilmark{1,15},
   Jane E.~Morrison\altaffilmark{16},
   Brian O'Sullivan\altaffilmark{17},
   K. M.~Pontoppidan\altaffilmark{1},
   M. W.~Regan\altaffilmark{1},
   M. E.~Ressler\altaffilmark{11},
   G. H.~Rieke\altaffilmark{16},
   Silvia~Scheithauer\altaffilmark{6},
   Helen Walker\altaffilmark{18},
   G. S.~Wright\altaffilmark{9},
   }
\altaffiltext{1}{Space Telescope Science Institute, 3700 San Martin
  Drive, Baltimore, MD, 21218, USA}
\altaffiltext{2}{Sterrenkundig Observatorium, Universiteit Gent,
              Gent, Belgium}
\altaffiltext{3}{Dublin Institute for Advanced Studies, School of Cosmic Physics, 31 Fitzwilliam Place, Dublin 2, Ireland}
\altaffiltext{4}{Centro de Astrobiolog\'ia (INTA-CSIC), Dpto Astrof\'isica, Carretera de Ajalvir, km 4, 28850 Torrej\'on de Ardoz, Madrid, Spain}
\altaffiltext{5}{Laboratoire AIM Paris-Saclay, CEA-IRFU/SAp, CNRS, Universit\'e Paris Diderot, F-91191 Gif-sur-Yvette, France}
\altaffiltext{6}{Max Planck Institute f\"ur Astronomy (MPIA), K\"onigstuhl 17, D-69117 Heidelberg, Germany}
\altaffiltext{7}{Deutsches Zentrum f\"ur Luft- und Raumfahrt (DLR), K\"onigswinterer Str. 522-524, 53227, Bonn, Germany}
\altaffiltext{8}{I. Physikalisches Institut. Universit\"at K\"oln, Z\"ulpicher STr. 77, 50937, K\"oln, Germany}
\altaffiltext{9}{UK Astronomy Technology Centre, Royal Observatory,
  Blackford Hill Edinburgh, EH9 3HJ, Scotland, United Kingdom}
\altaffiltext{10}{ETH Zurich, Institute for Astronomy, Wolfgang-Pauli-Str. 27, CH-8093 Zurich, Switzerland}
\altaffiltext{11}{Jet Propulsion Laboratory, California Institute of Technology, 4800 Oak Grove Drive,
  Pasadena, CA 91108, USA}
\altaffiltext{12}{Ames Research Center, M.S. 245-6, Moffett Field, CA 94035, USA}
\altaffiltext{13}{Leiden Observatory, Leiden University, PO Box 9513, 2300 RA, Leiden, The Netherlands.}
\altaffiltext{14}{SRON Netherlands Institute for Space Research, PO Box 800, 9700AV Groningen, The Netherlands}
\altaffiltext{15}{The Johns Hopkins University, Department of Physics and Astronomy, 366 Bloomberg Center, 3400 N. Charles Street, Baltimore, MD 21218, USA}
\altaffiltext{16}{Steward Observatory, University of Arizona, Tucson,
  AZ 85721, USA}
\altaffiltext{17}{Airbus Defence and Space, Gunnels Wood Road, Stevenage, Hertfordshire, SG1 2AS, UK}
\altaffiltext{18}{RALSpace, STFC, Rutherford Appleton Lab., Harwell, Oxford, Didcott OX11 0QX, UK}
\altaffiltext{}{}
\altaffiltext{}{}
\altaffiltext{}{}
\altaffiltext{}{}
\altaffiltext{}{}
\altaffiltext{}{}
\altaffiltext{}{}
\altaffiltext{}{}
\altaffiltext{}{}

\begin{abstract} 
We describe the operations concept and data reduction plan for the
Mid-Infrared Instrument (MIRI) for the James Webb Space Telescope
(JWST).  The overall JWST operations concept is to use Observation
Templates (OTs) to provide a straightforward and intuitive way for
users to specify observations.  MIRI has four OTs that correspond to
the four observing modes: 1.) Imaging, 2.) Coronagraphy, 3.) Low
Resolution Spectroscopy, and 4.) Medium Resolution Spectroscopy.  We
outline the user choices and expansion of these choices into
detailed instrument operations.  The data reduction plans for MIRI are
split into three stages, where the specificity of the reduction steps
to the observation type increases with stage.  The 
reduction starts with integration ramps: stage 1 yields uncalibrated
slope images; stage 2 calibrates the slope images; and then stage 3
combines multiple calibrated slope images into high level data
products (e.g. mosaics, spectral cubes, and extracted source
information).  Finally, we give examples of the data and data products that
will be derived from each of the four different OTs.
\end{abstract}


\section{Introduction}
\label{sec_intro}

The Mid-Infrared Instrument (MIRI) for the James Webb Space Telescope
(JWST) will be used for all observations with JWST from 5 to
28.5~\micron.  MIRI provides observing capabilities for imaging,
coronagraphy, low-resolution spectroscopy (LRS) from 5-12~\micron, and
integral field unit (IFU) medium-resolution spectroscopy (MRS).  This
paper provides details of the operational model and data reduction
plans.  Other papers in this series give the science motivation
\citep[][hereafter Paper I]{rieke2014a}; describe the overall design
and construction of the instrument \citep[][hereafter Paper
II]{wright2014}; and then individually focus on the Imager
\citep[][hereafter Paper III]{bouchet2014}, the LRS
\citep[][hereafter Paper IV]{kendrew2014}, the Coronagraphs
\citep[][hereafter Paper V]{boccaletti2014}, the MRS
\citep[][hereafter Paper VI]{wells2014}, the detectors
\citep[][hereafter Paper VII]{rieke2014b}, and the focal plane system
\citep[][hereafter Paper VIII]{ressler2014}; they are followed by
a description of the expected
sensitivity and performance \citep[][hereafter Paper IX]{glasse2014}.

This paper first describes how users will request observations with
MIRI by using Observing Templates (OTs) (\S\ref{sec_ots}).  The
instrument operations that will be driven by the user requests are
described in detail.  Then the plans for reduction of the MIRI data
are given in \S\ref{sec_data_red}.  While the data reduction plans are
expected to change at some level as the understanding of the MIRI
instrument evolves with continued ground-testing and on orbit
observations, the overall plan and many of the details will continue
to be as described in this paper.  Finally, we show examples of
observations for each of the four major MIRI Observation Templates
(\S\ref{sec_examples}).  The overall goal of this paper is
to provide an overview of the combined MIRI data acquisition and reduction
that are designed to produce high level data products for all MIRI
observations.  The OTs encode the best data acquisition practices as
determined from the instrument experts.   The data reduction plan
encodes the best practices for producing the highest level data
products.  This plan is mainly composed of an automated data reduction
pipeline with interactive tools provide for the steps that cannot be
automated.  

\section{Observing Templates}
\label{sec_ots}

The operational model for JWST is to describe the observations 
using Observation Templates (OTs).  The OTs provide a
straightforward interface for astronomers to specify the full
sequence of steps requested for their programs. 
The detailed instrument operational commands are
directly driven by choices provided in the OTs.  The 
OTs embed the knowledge of instrument experts,
effectively exporting their knowledge to all astronomers.  The
principles that drive the OT design include redundant observations
(e.g., multiple dithers), minimizing the use of the instrument
mechanisms (e.g., filter and grating wheels), and creating the minimal
set of options that accomplishes all expected MIRI science while
ensuring robust data reduction and a homogeneous and high-quality
archive.  This approach has been applied successfully already for the
{\it Infrared Space Observatory} \citep{kessler2003}, the {\it Spitzer
Space Telescope} \citep{werner2004}, and the {\it Herschel Space
Observatory} \citep{pilbratt2010}. In all three cases, OTs were found
both to make it straightforward to specify observations and also to
increase the probability of obtaining high quality results.

There are four MIRI OTs corresponding to the four main observing
modes.  These are Imaging, Coronagraphy, Low Resolution Spectroscopy,
and Medium Resolution Spectroscopy.  The basic parts of a MIRI OT are
slewing to the science target, guide star acquisition, target
acquisition (if needed), filter/grating moves, taking science
exposures, dithering, and mapping (if requested).  The full details of
the OTs are given in the MIRI Operations Concept
Document\footnote{available in its current version from
  http://www.stsci.edu/jwst/instruments/miri/docarchive/} and we 
only provide an overview of the OTs here.  While the exact details of
the OTs may change before and after the launch of JWST, the overall
design should not.

\subsection{Guide Star Acquisition}

An observational sequence begins with a telescope slew to the target position and 
acquiring guide star. A guide star acquisition is needed for all observations to provide
stable pointing.  The Fine Guidance Sensor (FGS) instrument
\citep{doyon2012} provides the tracking information using known point
sources that are mainly from the Guide Star Catalog
\citep{lasker2008}.  This catalog has an accuracy of
$\sim$$0.4\arcsec$ \citep{morrison2001}.  For any observations that
require placement of a source in an aperture with a higher accuracy,
dedicated target acquisition operations are used.  The relative astrometric
limiting accuracy of the FGS is expected to be $\sim$5 milli-arcsec
\citep{chayer2010} providing sufficient accuracy for all the MIRI
apertures using the target acquisition algorithms given in the next
subsection.  The overall telescope pointing jitter is expected to be
similar to the FGS relative astrometric accuracy \citep{hyde2004}.

\subsection{Target Acquisition}

Target acquisition (TA) is used to place sources accurately in
apertures or subarrays that are small when compared to the 3$\sigma$
accuracy of the guide stars.  The TA is done on-board the JWST
spacecraft and this places limitations on the complexity of the
algorithms used.  For the bright source TA, a clean image of the
source is created by taking 4 frames of data in a defined region of
the detector, ignoring the 1st frame to avoid detector reset
transients, making two estimates of the slope pixel-by-pixel by
differencing pairs of frames (i.e. 3-2 and 4-3), and taking the
minimum of the two estimates to reject cosmic rays
\citep{gordon2008a}.  The position of the source in the resulting
clean image is determined using a floating centroid algorithm
\citep{meixner2007}.  The on-board system uses the measurement of the
source position and the known central position of the aperture of
interest to perform a small move of the telescope to put the source in
the aperture.  For faint sources that require longer exposure times to
provide the necessary centroiding accuracy, a modified slope creation
algorithm is used and the exposure time needed is computed from the
flux of the source and the necessary centroiding accuracy
\citep{gordon2008b}.  Three of the standard MIRI imaging filters
(F560W, F1000W, and F1500W) and a dedicated neutral density filter
allow for TA to be done with blue or red sources and for a range of
brightnesses.  Filters with central wavelengths longer than
$\sim$15~\micron\ are not used for TA as the larger size of the
diffraction limited MIRI point spread function does not allow for high
enough centroiding accuracy \citep{gordon2008a}.

\subsection{Visits}

An observation consists of a series of observatory and instrument
commands.  To facilitate efficient scheduling, OT requests are split
into a set of Visits.  Visits are defined as uninterruptable blocks of
commands that carry out the user requested exposures.  An observation
can be broken into multiple Visits if a new guide star acquisition is
needed (e.g. due to a large requested change in the telescope
pointing) or if a different instrument mode is used (e.g. changing
from MIRI Imaging to MIRI Coronagraphy).  As a specific example, the
different tiles in a large mosaic are likely to be different Visits,
with small scale dithers and exposures with different filters all
taking place in a single Visit at each mosaic tile position.  Visits
are nominally defined such that they do not depend on each other,
allowing the scheduling system to interleave Visits from different OTs
and/or required instrument or observatory maintenance.

\subsection{Special Requirements}

There are cases where the ordering or timing of a set of Visits is
critical to carrying out the necessary observations.  Special
Requirements are used to indicate such requests to the schedule
system.  For example, coronagraphic observations often include
measurements of a reference PSF star as well as the target star.
These observations should be taken with back-to-back Visits with a
Special Requirement to indicate the necessary Visits should be taken
without any interruptions.

\subsection{Exposures}
\label{sec_exposures}

The MIRI instrument uses 1024$\times$1024 Si:As detector arrays,
described in more detail in Papers VII and VIII.  For the purposes of
MIRI operations, it is useful to know that the detectors are
non-destructively read out in equal intervals using four readout
channels, until they are reset.  A MIRI exposure can consist of
multiple integrations where an integration is defined as the time
between subsequent detector resets.  The time between successive reads
of a pixel is 2.775 s when reading the full detector in the FAST
readout mode and 27.75 s in the SLOW readout mode, so the
integration times are quantized in these units.  Readout patterns 
suitable to an observation type are specified in the appropriate OT.


\subsection{Imaging Observing Template}

The full details of the Imager design, build, and ground-testing are
given in Paper III. In summary, the MIRI Imager (MIRIM) has 9
wide-band filters with central wavelengths of 5.6, 7.7, 10.0, 11.3,
12.8, 15.0, 18.0, 21.0, and 25.5 $\micron$ and a plate scale of
0.11$\arcsec$ pix$^{-1}$. The MIRIM is critically sampled at 7.0
$\micron$ and has an unobstructed FULL array field-of-view (FOV) of
74\arcsec$\times$113\arcsec. In addition to the FULL array, MIRIM has
four other subarrays that can be used (BRIGHTSKY, SUB256, SUB128, and
SUB64); subarrays are required to observe bright targets (e.g. nearby,
transiting exo-planet host stars) and might be called upon as a
contingency should the telescope thermal emission be unexpectedly high
at wavelengths, $\lambda$ $>$ 20 $\micron$. The Imaging OT will be
used to request all imaging observations, including observations of
single fields, mosaicked fields, and non-contiguous fields.

The user options for the Imaging OT are summarized in
Table~\ref{tab_imaging_template}. First, the observer selects the
full array or a subarray and the dither pattern(s) to use. If the target requires
precision synoptic photometry using the SUB64 subarray (e.g.,
exo-planet transits), then a TA is required to place the target
accurately at the center of the subarray. In this case, the observer
must specify the TA filter and expected flux in the filter. If the
target does not require precision synoptic photometry, then the
observer selects a primary and/or a secondary dither pattern.

For full-array imaging, a primary dither pattern moves the target over a
significant fraction of the detector area\footnote{The maximum size of
the dither pattern that does not require re-acquisition of guide stars
is under evaluation.} to mitigate the effects of bad pixels and
provide data for self-calibration.  The secondary pattern moves the
target by a handful of pixels to improve sub-pixel sampling. The
dither pattern may be 'Cycling', 'Gaussian', or 'Reuleaux' and can be
scaled to have small, medium, or large overall scalings
\citep{chen2010}.  For subarray 
imaging, the dither options depend on the subarray selected so as not
to have excursions outside the imaging area. Once the imaging area
(array or subarray) and dithering pattern are selected, the observer
specifies the imaging filters and their corresponding exposure
times. All of the filters selected within a template will be executed
with a common subarray and dither pattern; therefore, the observer
must take care to ensure that the selected dither pattern addresses
their science goals. For example, the F560W filter is mildly
under-sampled with 1.7 pixels per FWHM and observers planning to use
PSF reconstruction algorithms should select a secondary dither
pattern.  However, an observer requesting F560W and F2550W
observations together may prefer not to select a secondary dither pattern if
PSF reconstruction is not a science driver for their program.

\begin{figure}[tbp]
\epsscale{0.8}
\plotone{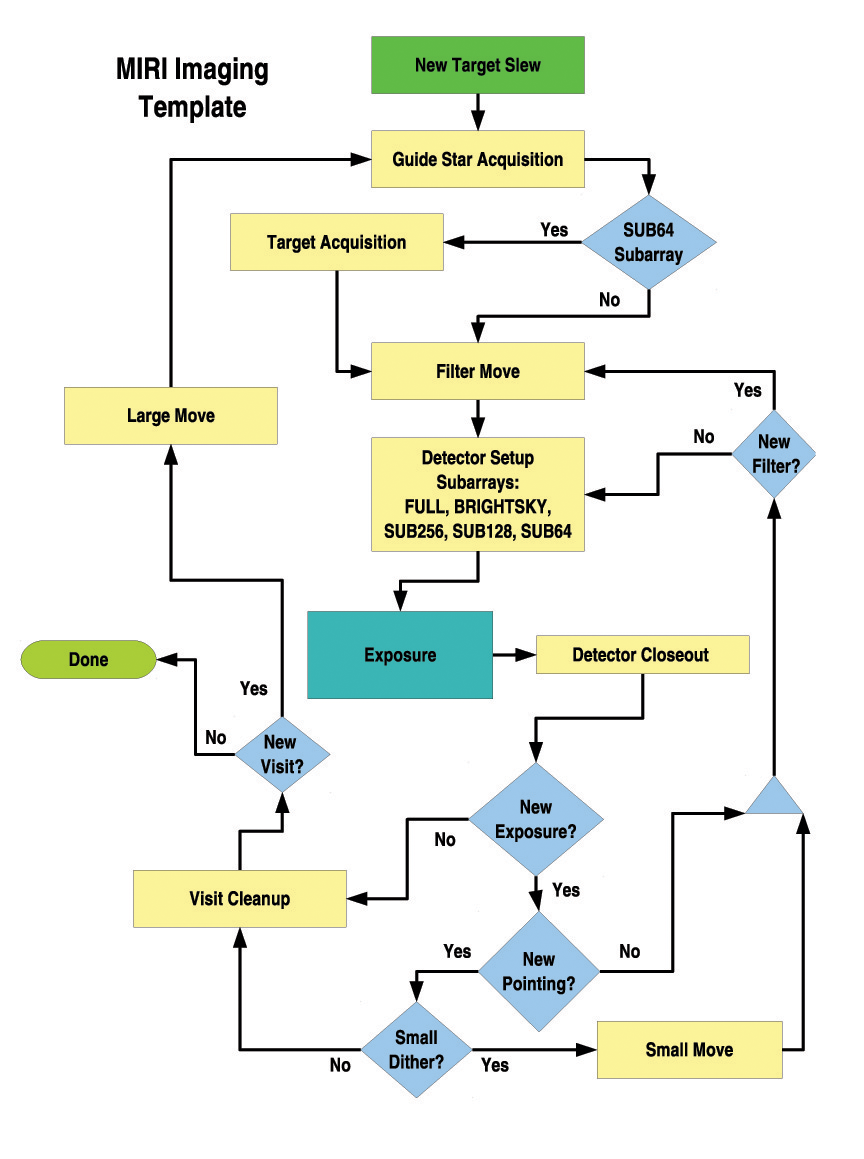}
\caption{The sequence of instrument and telescope operations in the Imaging OT are shown. The main overheads are shown in yellow boxes,
  decisions in blue boxes, and the actual observations in the  ``Exposure'' box.}
  \label{fig_imaging_template}
\end{figure}

The full sequence of instrument and telescope operations for the
Imaging OT is shown in Fig.~\ref{fig_imaging_template}. Note that
separate tiles within a mosaic may be considered large moves of the
telescope pointing and that such moves are scheduled in separate
Visits.

\subsection{Coronagraphy Observing Template}

MIRI has one Lyot and three 4 quadrant phase mask (4QPM)
coronagraphs.  The full details of the Coronagraphic design, build,
and ground-testing are given in Paper V.  The Lyot coronagraph is
designed for use with a wide-band filter with a central wavelength of
23~\micron.  The three 4QPM coronagraphs each are designed for a
specific wavelength and, thus, are used with medium-band filters with
central wavelengths of 10.65, 11.40, and 15.50~\micron.  The FOV of
the Lyot coronagraph is 30\arcsec$\times$30\arcsec, while the FOVs of
the 4QPM coronagraphs are all 24\arcsec$\times$24\arcsec.  Each
coronagraph has a dedicated subarray readout pattern. The Coronagraphy
OT will be used for all coronagraphic imaging observations.

The user options for the Coronagraphy OT are summarized in
Table~\ref{tab_coronagraphy_template}.  The coronagraph to be used is
automatically specified by choosing its associated filter.  The user
must specify the exposure times in each of the requested
coronagraphs. All coronagraphic observations require TAs to center the
source precisely; the FND filter is a neutral density filter allowing
TA on very bright targets.  The Lyot and 4QPM coronagraphs have
different TA procedures customized for the different coronagraphic
types; thus, the user must specify the TA filter and expected flux in
the filter for each coronagraph.

Most coronagraphic observations will require reference PSF
observations, as the JWST primary mirrors are re-phased periodically
and the wavefront errors will change between phasings.  The user will
specify the same information for the reference source as for the
target, except for the choice of coronagraphs.  For observations
focusing on detecting nearby compact sources, angular differential
imaging \citep[ADI,][]{marois2006} can be used, where exposures of the
source are taken with different telescope roll angles and then
subtracted.  Such observations can be specified with a Special
Requirement.  Due to the need to shade the JWST telescope from the
Sun, instantaneous roll angles are limited to $\pm 5^\circ$; ADI
techniques used must be consistent with this limitation.

\begin{figure}[tbp]
\epsscale{0.8}
\plotone{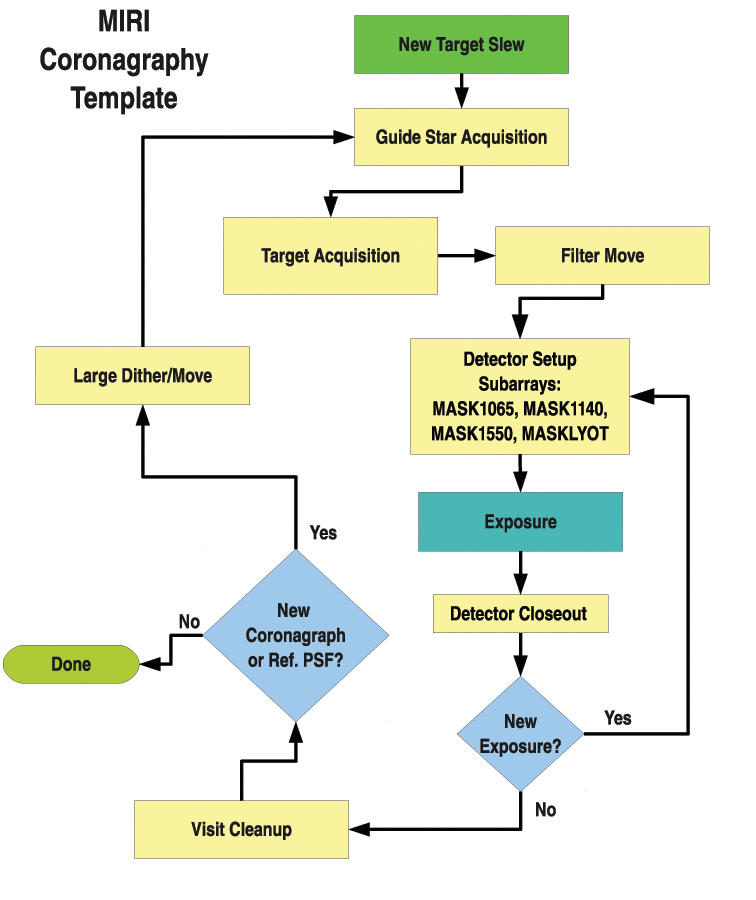}
\caption{The sequence of instrument and telescope operations in the Coronagraphy OT are shown.  The main overheads are shown in yellow boxes,
  decisions in blue boxes, and the actual observations in the  ``Exposure'' box.}
  \label{fig_coronagraphy_template}
\end{figure}

The full sequence of Coronagraphy OT instrument and telescope
operations is shown in Fig.~\ref{fig_coronagraphy_template}.

\subsection{Low Resolution Spectroscopy Observing Template}

The Low Resolution Spectrograph (LRS) provides the capability to take
5-12~\micron\ spectroscopy with a resolution of $\sim$100, using a double prism
mounted in the MIRIM filter wheel. The full details of
the LRS design, build, and ground-testing are given in Paper IV. The
LRS OT will be used to define observations of single sources, either
point or extended.  The majority of LRS observations are expected to use the $0.51\arcsec \times 4.7\arcsec$ slit that includes
a filter to block light below 5~\micron.  For observations desiring
the highest accuracy for bright variable sources (e.g. exoplanet
transits), there is a slitless mode that avoids variable slit
transmission losses.  The slitless mode uses the SLITLESSPRISM
subarray to allow bright sources to be observed without saturation.

The user options for the LRS OT are summarized in
Table~\ref{tab_lrs_template}.  LRS observations of point sources will
require a TA to place the source accurately to ensure a good
wavelength calibration.  This TA will be done using the science
target.  The user picks one of four TA filters and provides an
estimate of the flux in this filter.  The slit type will be specified
as slit or slitless and this decision will automatically select the
FULL or SLITLESSPRISM subarray as appropriate.  The dither type can be
'None' (e.g. for exoplanet observations), 'Point' where a point source
is dithered between positions that are located 1/3 and 2/3 along the
slit direction, or 'Extended' where the source is chopped between the
center of the slit and a region well outside the slit \citep{chen2009}.  Both the
'Point' and 'Extended' dither patterns provide data to allow the
removal of the background.  Finally, the user will input the requested
exposure time per dither position.

\begin{figure}[tbp]
\epsscale{0.9}
\plotone{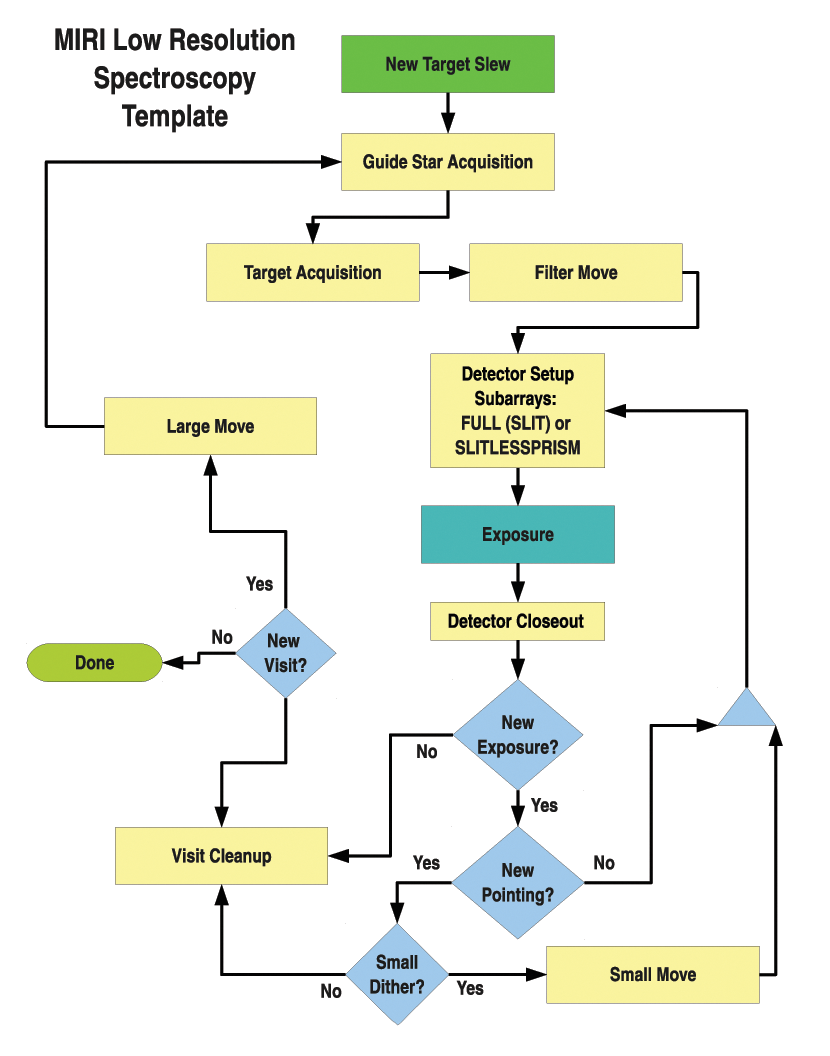}
\caption{The sequence of instrument and telescope operations in the LRS OT are shown.  The main overheads are shown in yellow boxes,
  decisions in blue boxes, and the actual observations in the   ``Exposure'' box.}
  \label{fig_lrs_template}
\end{figure}

The full sequence of LRS OT instrument and telescope operations is
shown in Fig.~\ref{fig_lrs_template}.

\subsection{Medium Resolution Spectroscopy Observing Template}

The Medium Resolution Spectroscopy (MRS) has four Integral Field Units
(IFUs) that feed medium resolution spectrometers operating from 5 to 28.5~\micron\ with a
resolution of $\sim 3000$.  The full details of
the MRS design, build, and ground-testing are given in Paper VI.  The
IFUs are nested, resulting in simultaneous observations in all four
spectrometer channels.  While this imposes the limitation that the
exposures times are the same in all four channels, it is possible to
have different length integrations (see $\S$\ref{sec_exposures})
between the two short wavelength 
channel (imaged onto a short wavelength optimized Si:As detector)
and two long wavelength channels (imaged onto a different long
wavelength optimized Si:As detector).  Each
IFU has a different FOV ranging from $\sim$3\farcs7$\times$3\farcs7\
at the shortest wavelengths to $\sim$7\farcs7$\times$7\farcs7\ at the
longest wavelengths.  Thus, the slice widths and detector plate scales
vary with IFU to match the varying diffraction limited PSF of JWST (as
discussed in detail in
Paper VI).

The MRS OT will be used to take observations of individual and
extended sources.  The wavelength coverage of each channel of the MRS
for a single grating setting is approximately 1/3 of the total
wavelength range.  Thus, three grating settings (A, B, \& C) are required to
obtain a full 5-28.5~\micron\ spectrum of a source.  Observations with
a single grating setting will produce 4 disjoint spectral segments,
one from each IFU.

The user options for the MRS OT are summarized in
Table~\ref{tab_mrs_template}.  A TA will be needed for most MRS
observations, requiring the user to specify the TA filter and expected
flux in this filter.  The user picks whether 1, 2, or all 3 settings
are to be utilized.  The dither patterns possible are none, 4-point, 6-point, or cycling \citep{chen2012}.  The four
IFU slice widths and pixel sizes are optimized to be appropriately
sub-sampled by a single 2-point dither pattern for all four IFU
channels. Specifically, an offset in the cross-slice direction of
0\farcs97 will sub-sample each slice by half a slice width plus an
integer number of slices. This is important for the shorter wavelength
IFUs as they are undersampled in the cross slice
direction. Simultaneously an offset in the along-slice direction by
7.5 pixels (1\farcs47) will subsample the pixels.  To mitigate bad
pixels, the two point pattern is repeated twice, creating an effective
4-point dither pattern, appropriate for extended sources. For compact
sources, where better sampling is needed, a finer 6-point dither is
available. Finally, for high-background regions, a self-calibration
cycling pattern can be used.

\begin{figure}[tbp]
\epsscale{0.9}
\plotone{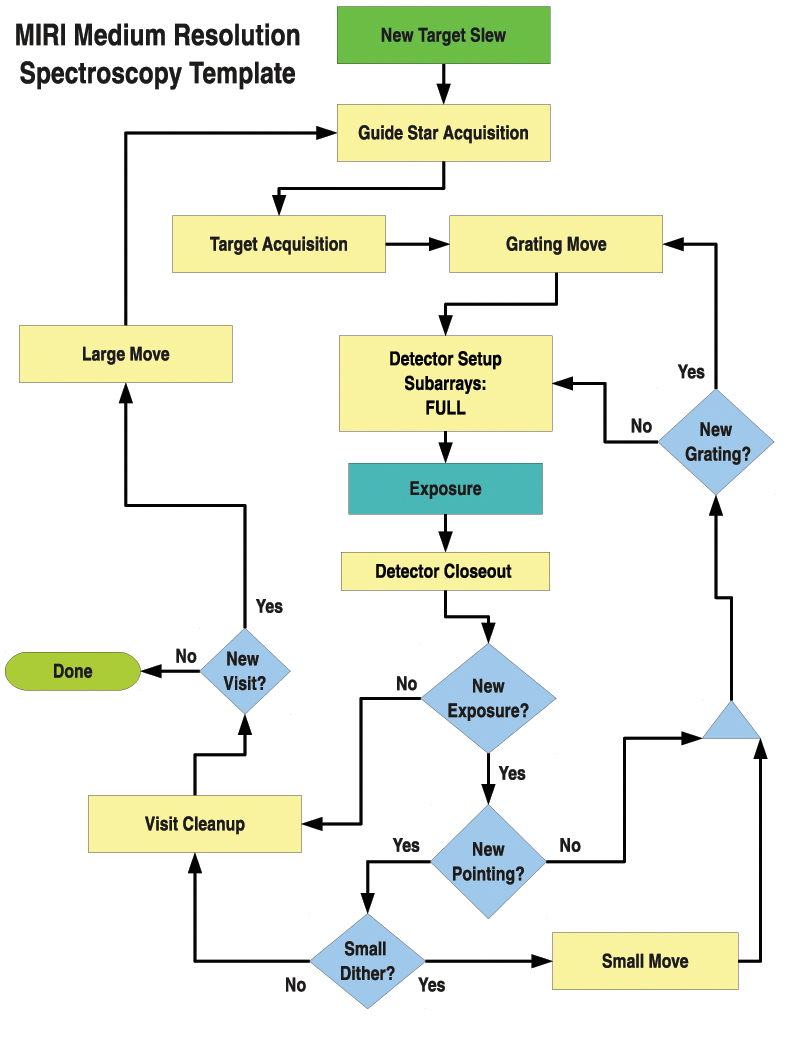}
\caption{The sequence of instrument and telescope operations in the MRS OT are shown.  The main overheads are shown in yellow boxes,
  decisions in blue boxes, and the actual observations in the   ``Exposure'' box.}
  \label{fig_mrs_template}
\end{figure}

The full sequence of MRS OT instrument and telescope operations is
shown in Fig.~\ref{fig_mrs_template}.

\section{Data Reduction Plan}
\label{sec_data_red}

\begin{figure*}[tbp]
\epsscale{1.0}
\plotone{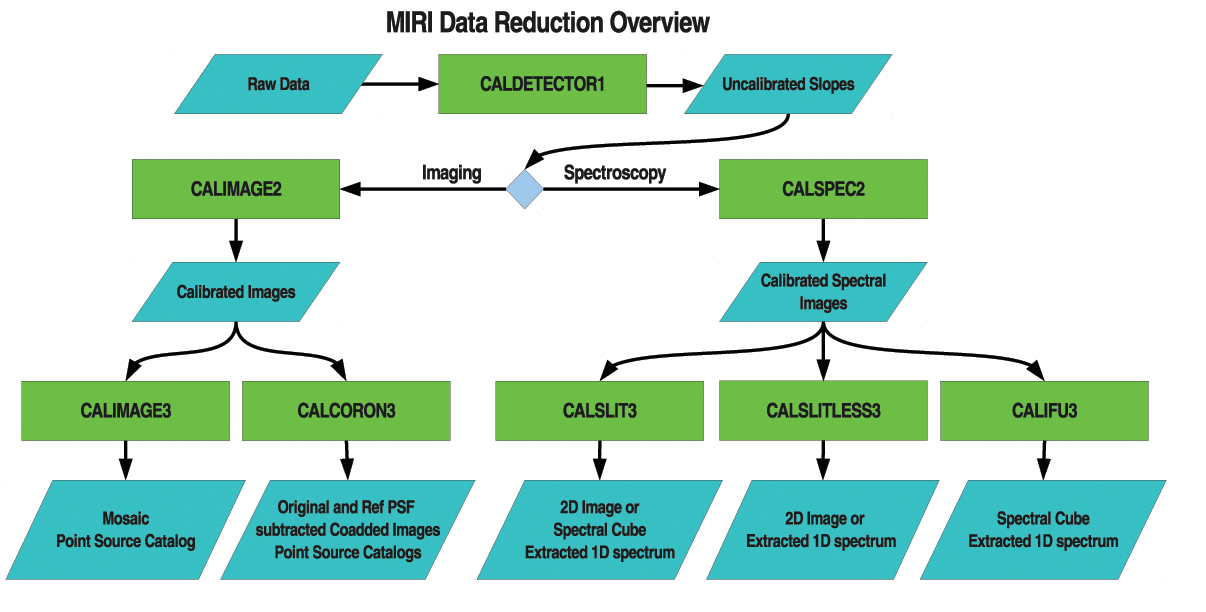}
\caption{The overview of the MIRI data reduction plan.}
  \label{fig_cal_overview}
\end{figure*}

The MIRI data reduction plan is based on the extensive testing of the
instrument (Paper II), previous work on reducing data from similar
instruments and telescopes, and on a series of test campaigns being
conducted at JPL using flight-clone detectors and a flight-like data
chain.  The data reduction is split into 3 stages; an overview is
presented in Fig.~\ref{fig_cal_overview}.  The 1st stage
(CALDETECTOR1) processes the raw data, composed of non-destructively
read ramps for each pixel, into an uncalibrated slope image.  The 2nd
stage (CALIMAGE2 and CALSPEC2) calibrates each slope image.  The 3rd
stage (CALIMAGE3, CALCORON3, CALSLIT3, CALSLITLESS3, and CALIFU3) uses
the set of slope images taken for an observation and processes them to
produce the final data products.  The data reduction plan is not restricted to single instances of an
OT; observations from multiple instances of OTs can be
combined together in the later pipeline stages.  Where possible
decision points potentially exist (e.g. point or extended source
spectral extractions), the philosophy will be to process the
observations with both assumptions and provide both to the end user
who can then decide which is best for their analysis.
A description of each
of these stages is given in the following subsections.  Some aspects
of the data reduction algorithms are expected to change in the coming
years as the understanding of the MIRI detectors and instrument
improves.  However, as this evolution will not be radical, the current
plans should continue to provide a valid overview.

\subsection{Pipeline Stage One}
\label{sec_stage_one}

\subsubsection{CALDETECTOR1}

\begin{figure}[tbp]
\epsscale{1.0}
\plotone{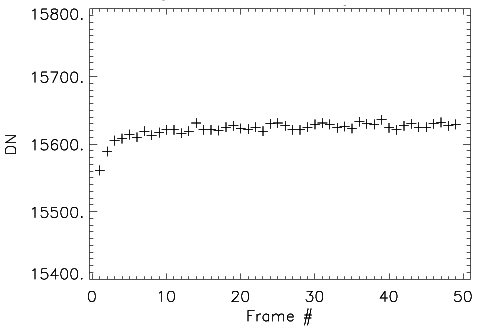}
\caption{A data ramp for a pixel (in DN) is plotted for a dark
exposure. The effect of the reset anomaly is shown by the initial
frames being offset from their expected values.
  \label{fig_reset}}
\end{figure}

\begin{figure}[tbp]
\epsscale{1.0}
\plotone{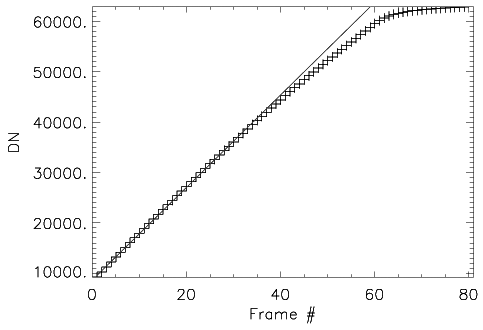}
\caption{An example of a data ramp for a pixel that saturates. A
linear fit to
the first 35 frames is shown in blue. The reduction in the responsivity 
with increasing signal is shown by the departure in the frame values from the
linear line.
  \label{fig_linearity}}
\end{figure}

The first stage in the pipeline, CALDETECTOR1, is to measure the slope
(DN/s units) from the non-destructively read data ramps for each
pixel. The data ramps show a number of non-ideal effects common to
Si:As devices, including non-linearity, reset anomaly, and latent
images, as well as slow drifts in the slopes with amplitudes roughly
proportional to the signal level and slow drifts in the zero point of
the ramp (see Papers VII and VIII). Here we show two examples.  The
first is the reset anomaly as shown in Fig.~\ref{fig_reset}, where the
first few frames in a dark exposure are lower than expected.  The
second is non-linearity as shown in Fig.~\ref{fig_linearity}, which
shows the data ramp for a brightly illuminated pixel that saturates
around frame 60. The data values are plotted as plus symbols and a
linear fit using the first thirty-five frames is overplotted in
blue. The classical non-linearity response of the pixels with
increasing signal is shown by the departure of the frame values from
the linear line.

Taken all together, these non-ideal detector characteristics pose a
challenge to accurate data reduction, since the full suite of issues
can be influenced by a large variety of parameters including
illumination history, illumination brightness, and time since last
exposure. The approach we have adopted is to assume the different
detector effects do not depend on each other (e.g. they are
orthogonal).  This is a standard assumption in most data reduction
algorithms, as applied for instruments operating at wavelengths from
the ultraviolet to the far-infrared.  For example, this assumption
has been used successfully for the data reduction of the 24~$\micron$
observations taken with the Multiband Imaging Photometer for Spitzer
\citep{gordon2005}.
The ordering of the steps that correct the non-ideal characteristics
is important and is done in the reverse of the order that they happen
in the detector and electronics.  The current steps include 1)
rejecting saturated data or bad data based on a predefined bad pixel
mask, 2) removing common noise components, 3) correcting for anomalies
in the initial frames in an integration caused by the reset, 4)
correcting for the contamination of the data by a source from the
previous exposure (commonly referred to as persistence), 5) correcting
for the non-linearity of the ramps caused by the debiasing of the
detectors, 6) subtracting the contribution of the dark current from the
ramps, 7) detecting jumps in the ramps caused by cosmic rays and noise
spikes \citep{anderson2011} and, finally, 8) determining the slope of
each pixel by fitting 
line segments to each data ramp. We expect additional steps may be
introduced as our understanding of the detectors improves.  The
ordering of the steps may also change.

\subsection{Pipeline Stage Two}
\label{sec_stage_two}

The 2nd stage of the pipeline focuses on corrections for pixel- and
time-dependent effects on the individual images.  Specifically, the
corrections are performed on the slope images created by CALDETECTOR1
and produce calibrated images in physical units (e.g., MJy/sr and
wavelengths in $\micron$).  The reduction splits into two branches
with the CALIMAGE2 branch for imaging and coronagraphy and the
CALSPEC2 branch for the LRS and MRS spectroscopy.

\subsubsection{Common Steps}
\label{sec_stage_two_com}

The first two steps in both branches are the same and are:

\medskip

\noindent {\it ~ Residual Persistence Correction:} The main correction
for persistent images (also called latents) is done in CALDETECTOR1,
but residual persistence will be corrected using measurements of
images taken prior to the current image.  The persistence signal to
subtract will be constructed from the previous images and a
persistence model with decay amplitudes and time constants calibrated
using dedicated persistence calibration observations.

\medskip
\noindent {\it ~ Flat-Fielding:} The variations in sensitivity from
pixel-to-pixel will be corrected using flat field images.  For MIRIM,
it should be possible to generate flat fields from sky signals and
standard self-calibration reduction procedures. The uniformity of the
Imager response will be determined by dedicated observations of point
sources. For the MRS, flat field measurements will utilize the
internal flat field source (small spatial scale variations) and
external point sources (large spatial scale variations).  The flat
fields used will be specific to the OT used.

\subsubsection{CALIMAGE2}

The stage 2 portion of the pipeline for imaging (including
coronagraphy) is called CALIMAGE2.  The CALIMAGE2 set of algorithms
primarily processes single exposure slope images without extensive
reference to other slope images taken as part of the OT.  The data
reduction steps in this stage are similar for the Imaging and
Coronagraphic observations, with the main differences being in the
calibration files and minor differences in the algorithms for some of
the steps.  The CALIMAGE2 steps beyond the two already given in
\S\ref{sec_stage_two_com} are:

\medskip
\noindent

{\it Spatial Mapping:} The mapping from pixel to spatial coordinates
for each slope image will be computed and associated with each
exposure.  This provides the information needed to map each pixel to
the correct position on the sky.  These mappings are mainly based on
extensive ground-test distortion measurements and will be confirmed
with on-orbit observations.

\medskip
\noindent

{\it Absolute Photometric Calibration:} The conversion from instrument
units (DN/s) to physical units (MJy/sr) is done by multiplying by the
appropriate calibration factor.  The calibration factors will be
derived using Imaging and Coronagraphic observations of flux
calibration stars \citep{gordon2009, gorboh2009}.

\subsubsection{CALSPEC2}

The stage 2 portion of the pipeline for spectroscopy is called
CALSPEC2.  The goal of this stage is to remove additional non-ideal
instrumental effects from the slope images for spectroscopic
observations.  The CALSPEC2 set of algorithms primarily processes
single exposure slope images without extensive reference to other
slope images taken as part of the OT.  The data reduction steps in
this stage are similar for the LRS and MRS observations, with the main
difference being the calibration files and minor differences in the
algorithms for some of the steps.  The CALSPEC2 steps beyond the two
already given in \S\ref{sec_stage_two_com} are given below in the
current planned order.

\medskip
\noindent

{\it Straylight Correction:} Stray light that contaminates the
observed spectra is corrected by subtracting a model of the scattered
light.  The model for the LRS observations takes into account the
light scattered and diffracted from other regions of the detector.
For example, the LRS slit observations are measured in a detector
region between the regions where the coronagraphic and imaging
observations are taken.  For the MRS observations, the model will
include the small fraction of the light that is scattered off the
optical surfaces and reaches the detectors. Correcting for the
extended detector response artifacts between 5 and 8 $\mu$m (Paper
VII) is under consideration for this stage.

\medskip
\noindent

{\it MRS Fringing:} The spectral fringing in the MRS observations
(Paper VI) will be corrected using a model of the fringes based on
ground-testing data and refined using on-orbit data to generate a
fringe flat field\footnote{An interactive tool will also be provided
to assist users in improving the fringe correction beyond that from
the pipeline.}.

\medskip
\noindent

{\it Spatial and Spectral Mapping: } The mapping from pixel to spatial
and spectral coordinates for each slope image will be computed and
associated with each exposure.  This provides the information needed
to map each pixel to the correct position on the sky and in wavelength
space.  These mappings are mainly based on extensive ground-test
results and will be confirmed with on-orbit observations.  Any
variations in the expected centering of point sources in the slit
(LRS) or slices (MRS) will be included in the mapping solution.

\medskip
\noindent

{\it Absolute Spectrophotometric Calibration:} The conversion from
instrument units (DN/s) to physical units (MJy/sr) is done by first
multiplying by the appropriate Spectral Response Functions (RSRFs)
that provide the calibration factor as a function of wavelength
(Papers IV and VI).  Initial RSRFs have been obtained in ground test
and will be refined using LRS and MRS observations of stars. The
absolute calibration will be based on standard stars \citep{gordon2009, gorboh2009}.

\medskip
\noindent

{\it LRS Compact Source Background Subtraction:} The background
subtraction for LRS exposures for compact sources is straightforward.
Compact sources will be nodded/dithered between two different
positions in the slit.  The background can either be removed by
subtracting the two different nod positions, or by using a polynomial
fit to regions in the slit with minimal contributions from the source,
or by a combination of both approaches.  For extended sources, a
separate background observation will be used to subtract the
background.

\medskip
\noindent

{\it LRS Compact Source Extraction:} For sources more compact than the
LRS slit length, spectra will be extracted using fixed apertures and
optimal extraction using a Point Response Function (PRF). The PRF is a
slit position dependent table that combines the information on the
PSF, the detector sampling, fringing, and the intra-pixel sensitivity
variations.  For the fixed aperture extraction, fringing and
intra-pixel corrections are applied to the extracted 1D spectra.  For
optimal extraction, these corrections are included in the PRF to
ensure that low signal-to-noise pixels get low weighting.
Observations of point sources through a slit can show variations due
to variations in the telescope pointing (e.g. jitter).  If the
variations are significant, a delta throughput correction will be
applied using time resolved telescope pointing information.

\subsection{Pipeline Stage Three}

The goal of the 3rd stage of the pipeline is to process multiple
exposures and extract information for specific sources to produce high
quality final data products that are ready for science analysis.  This
includes combining multiple exposures and extracting source
information from the individual and combined exposures.  The reduction
splits into five branches at this stage. Each of these branches
(CALIMAGE3, CALCORON3, CALSLIT3, CALSLITLESS3, and CALIFU3) addresses
issues that are specific to the details of the observing mode.  Two
important concepts for these branches are:

\medskip
\noindent

{\it JWST Background: } One of the critical steps for most MIRI
observations at this stage is to account for the background.  The JWST
background at MIRI wavelengths will be significant and it is critical
to remove and/or model this background to produce high quality
reductions.  This background is from astrophysical (scattering and
emission from Zodical and Milky Way dust) and spacecraft (telescope
and sunshade) sources, with the former dominating at short wavelengths
and the latter dominating at long wavelengths.

\medskip
\noindent

{\it Spectral Cubes:} One data product that has significant utility
for inspection and data analysis of spectral observations, mainly for
the MRS IFU spectrometer but also for LRS mapping spectra, is a
spectral cube. A spectral cube has the spectral data reprojected from
the 2D detector plane to a 3D cube with two spatial and one spectral
dimension.  The cube is constructed such that each spatial plane is a
monochromatic image of the source.  The observations that are used to
construct the cube can be a single exposure or multiple exposures with
different gratings or dither positions.

\subsubsection{CALIMAGE3}

The 3rd stage of the pipeline for Imaging is called CALIMAGE3.  The
goal of CALIMAGE3 is the processing of multiple exposures to produce
high level imaging products.  These data products are mosaics of
multiple individual exposures and catalogs of point sources.  The
steps in the current planned order are:

\medskip
\noindent

{\it Background Matching:} The background may change between imaging
exposures (particularly if they are not contiguous) due to changes in
the telescope emission or zodiacal light.  The overlapping regions
between successive images will be used to correct for differences in
the background between images.

\medskip
\noindent

{\it Delta Cosmic Ray Detection: } The overlapping regions between any
images in the region will be used to detect and reject from further
processing residual cosmic rays that were not detected in the ramp
processing (see \S\ref{sec_stage_one}).

\medskip
\noindent

{\it Self Calibration: } For observations with sufficient redundancy
and appropriate dither patterns, self calibration algorithms
\citep{arendt2000, fixsen2000} will be used to test the dark and flat
field frames. If improvements are indicated, the same algorithms will
be employed to produce "delta-dark" and "delta-flat" corrections that
are correlated with instrumental (not sky) coordinates.  The
"delta-dark" will include the residual additive instrument signatures
(dark and background variations).  The "delta-flat" will include
multiplicative instrumental signatures (flat field variations).

\medskip
\noindent

{\it Source Extraction: } Standard PSF fitting algorithms will be used
to produce catalogs of point sources to provide at least an initial
view of the point sources in the images.  The tool that does the
source extraction will be available to the user to allow for user
interaction to create improved source catalogs.

\subsubsection{CALCORON3}

The 3rd stage of the pipeline for Coronagraphy is called CALCORON3.
The goal of CALCORON3 is to combine multiple images of the source and
reference star and produce images that suppress/remove the central
source PSF as much as possible.  The steps in the current planned
order are:

\medskip
\noindent

{\it Image Coaddition:} Multiple images of a source are coadded to
produce a high S/N image.  The coaddition is done in pixel coordinates
as dithering is not possible for coronagraphhic observations.

\medskip
\noindent

{\it Reference PSF Subtraction:} The reference PSF image is subtracted
from the science target image.  This suppresses the residual speckles
and significantly improves the contrast and signal-to-noise of faint
sources near the central source.  This can be a simple subtraction of
a directly observed reference star or a composite reference PSF
created from a library of observed reference stars.  This composite
reference PSF would be created using the LOCI
\citep[e.g.,][]{lafreniere2007}, KLIP \citep[e.g.,][]{soummer2012},
ADI \citep[e.g.,][]{lafreniere2007, janson2008}, or similar
algorithms.  The composite reference PSF algorithms directly handle
the variations in the source PSF due to changes in the telescope
segment phasing and pointing jitter.

\subsubsection{CALSLIT3}

The 3rd stage of the pipeline for LRS spectroscopy with the LRS slit
is called CALSLIT3.  The goal of CALSLIT3 is the processing of
multiple exposures to produce high level spectral products.  The steps
are given below, with the differences planned between the extraction
of point versus extended sources highlighted.

\medskip
\noindent

{\it Compact Sources Combined Spectra:} For point sources, the spectra
extracted from individual exposures will be coadded to produce a final
spectrum of the source. The background subtraction for compact sources
is already performed in CALSPEC2.

\medskip
\noindent

{\it Extended Source Spectral Cubes:} For sources that are extended
beyond the slit and have been taken with mapping dither patterns, a
spectral cube can be constructed using the known transformation
between the detector 2D coordinates to 3D spectral cube coordinates.
The tool described in the CALIFU3 section below for extraction of
extended sources will be used to extract extended source information
for LRS mapping observations.

\subsubsection{CALSLITLESS3}

The 3rd stage of the pipeline for LRS slitless spectroscopy is called
CALSLITLESS3.  The slitless mode is designed for high precision
observations of variable sources (e.g., exoplanet transients).  As the
goal of this mode is extreme relative photometric precision, the
possible changes in the telescope PSF, fringes, and intra-pixel
sensitivity among different exposures during the observation need to
be corrected.  These corrections will be derived from the observations
themselves including the use of a detailed model of the PSF
variations.

\subsubsection{CALIFU3}

The 3rd stage of the pipeline for MRS spectroscopy is called CALIFU3
and has the goal of processing multiple exposures to produce final,
high-level products.  The steps include background subtraction,
creation of spectral cubes, and extraction of sources from the
individual exposures and spectral cubes.  The CALIFU3 steps in the
current planned order are given below.

\medskip
\noindent

{\it Background Subtraction:} For compact sources with sizes smaller
than the offsets between dither positions, the background (and dark current) can be
subtracted from one dither position using the other dither positions.
For extended sources, either a dedicated off-source observation can be
used for background subtraction or, if such observations are not
possible, a self-calibration algorithm \citep{arendt2000, fixsen2000}
can be used.

\medskip
\noindent

{\it Spectral Cube Construction:} The information required to map the
detector coordinates to the 3D cube is derived from extensive
ground-testing and confirmed on-orbit with mapping observations of a
point source.  Such cubes will be constructed for individual exposures
as well as for the combinations of exposures dithered and with different
gratings/channels \citep[e.g.,][]{glauser2010}.

\medskip
\noindent

{\it Spectral Extraction:} The MRS spectra will be extracted from the
individual exposures and spectral cubes.  Default extractions assuming
a point source and a fully extended source will be provided from the
pipeline.  A tool that requires user interaction will be provided to
allow more sophisticated extractions.  One example of such an
extraction would be where the variance of the output spectrum is
minimized \citep{horne1986, marsh1989, robertson1986}.

\section{Data and Data Products Examples}
\label{sec_examples}

\begin{figure*}[tbp]
\epsscale{1.05}
\plotone{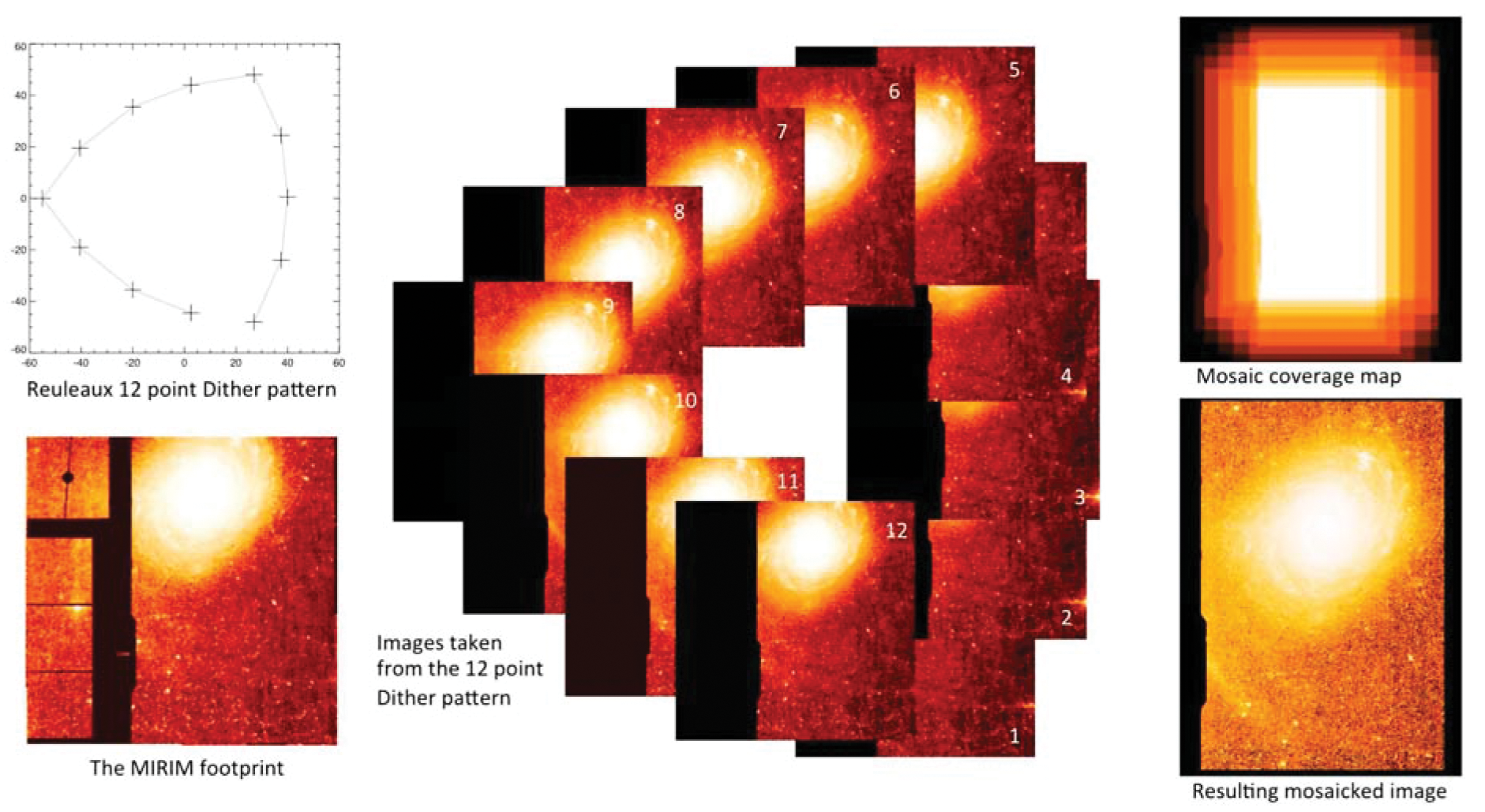}
\caption{An example of imaging observations and data reduction. 
Twelve images with the 12 point Reuleaux dither pattern are
shown separately and combined into a mosaic.  The
observations taken through the Coronagraphs are not used in the
production of the final mosaic for this simulation.}
\label{fig_imag_example}
\end{figure*}

Imaging observations of an galaxy are shown in
Fig.~\ref{fig_imag_example} to illustrate a typical set of Imager data
and the corresponding data reduction.  The observation sequence includes 12 individual
exposures, each taken at a different position in a Reuleaux dither
pattern.  The full MIRIM footprint is shown in the lower
left illustrating that 
observations are obtained in the Imager, Corongraphic, and LRS
FOVs.  For simplicity, only the data taken in the Imager FOV is used
to create the final mosaic.

\begin{figure*}[tbp]
\epsscale{1.05}
\includegraphics[scale=0.45,angle=0]{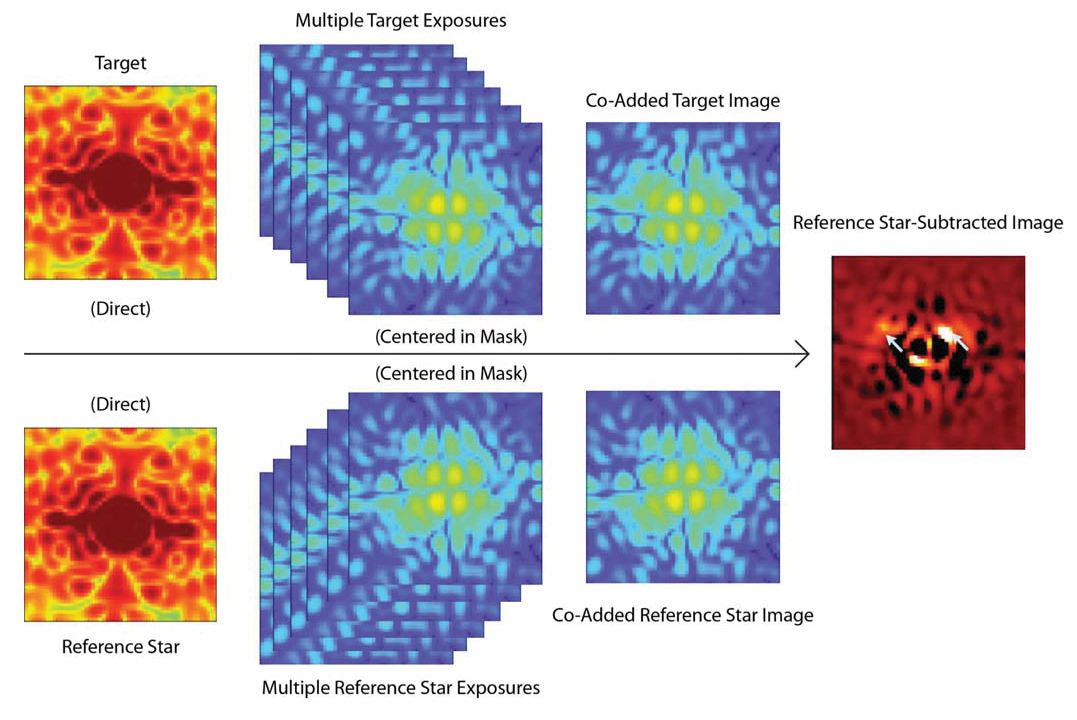}
\caption{An example of Coronagraphic observations and data reduction.
  Multiple images of the target and reference stars are combined to
yield a high quality subtraction of the reference star PSF from that
of the target star, revealing faint companions to it.}
\label{fig_coron_example}
\end{figure*}

An example of a typical MIRI 4QPM Coronagraphy observing and data
reduction sequence is shown in Fig.~\ref{fig_coron_example}.  A target
object and a reference star are each observed separately with the
coronagraph, and then later processed to form the final reference
star-subtracted image of the target.  The top and bottom sequences
represent the separate observations of the target and the reference
star.  The sequence begins with an observation of the object outside
the coronagraph through a TA filter.  Once the centroid of the object
is obtained, the telescope is offset to place the object in the center
of the coronagraph.  The correct coronagraphic mask/filter is
selected, and exposures are obtained.  Typically, multiple exposures
of an object are performed to achieve the required signal-to-noise in
the scientific region of interest, or to achieve the required
signal-to-noise in the diffraction pattern, or both.  The multiple
exposures are then co-added to form a single image of the target
object and the reference star.  The final image is formed by
subtracting the co-added reference star image from the co-added target
star image.

\begin{figure}[tbp]
\epsscale{1.05}
\plotone{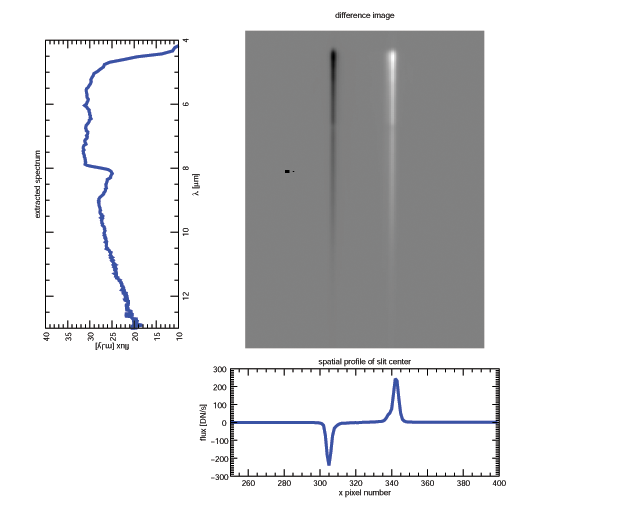}
\caption{An example of LRS data reduction for a point source taken in
two different positions in the slit.  The difference between the two
observations removes the background as shown in the grayscale image.
A spatial cut is shown in the lower plot and the extracted spectrum in
the left plot.  The flux in the spectrum was converted into physical units by using the
Relative Spectral Response Function shown in Paper IV.}
\label{fig_lrs_example}
\end{figure}

For LRS observations, a common observation set will be nodded
observations of a single source.  The nods are done at two
positions in the slit; the source spectrum can then be extracted
directly from the difference image.  An example of this type of data,
difference image, and extracted spectrum is shown in
Fig.~\ref{fig_lrs_example}.

\begin{figure*}[tbp]
\epsscale{1.0}
\plotone{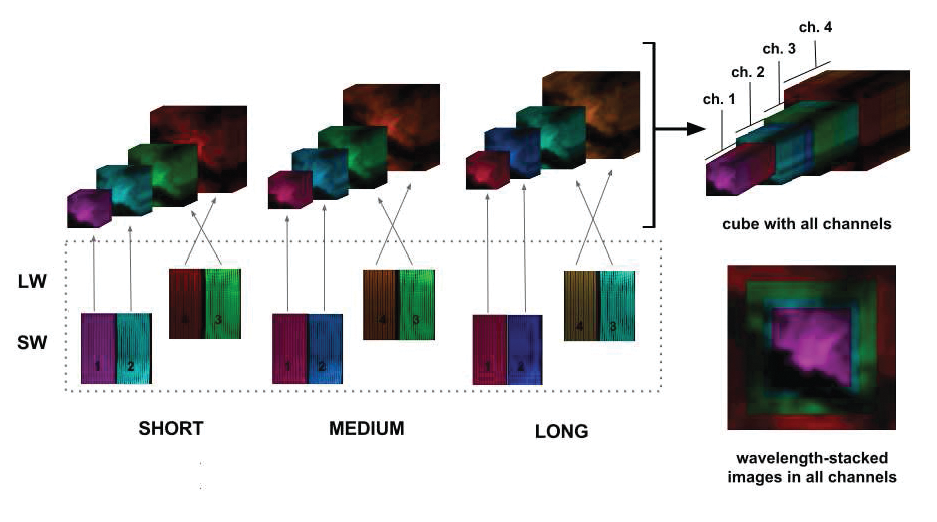}
\caption{An example of the MRS observations that would go into
constructing a full spectral cube of a source from 5-28.5~\micron.
The SW and LW rows give examples of the short and long wavelength
detector images.  The LONG, MEDIUM, and SHORT columns give the A, B,
and C grating
settings to obtain full spectral coverage in each of the four IFUs.
The detector exposures are color coded by IFU and correspond to the
individual IFU spectral cubes (left top row) Finally, the top row,
right has a full, merged spectral cube of all four IFUs and the bottom
row, right gives a monochromatic image from each IFU illustrating the
changing FOV from the shortest to longest wavelength
IFU. \label{fig_mrs_example}}
\end{figure*}

Fig.~\ref{fig_mrs_example} gives a graphical representation of the
construction of the full MRS spectral cube starting with individual
detector images.  An extended source is observed with the three
grating settings to obtain the full spectral
coverage in all four IFUs.  Such observations will often be taken with
4 dither positions at each grating position, resulting in 24 detector
images that are then processed to produce the full wavelength range
spectral cube.  This full spectral cube can be pictured to be composed of
four spectral cubes, one for each IFU.

\section{Summary}

The current plans for the operations strategy and data reduction for
MIRI have been described.  The operations strategy centers around the
use of Observing Templates to allow astronomers to give only the
necessary level of detail to carry out their desired observations.
The Observing Templates for MIRI are Imaging, Coronagraphy, Low
Resolution Spectroscopy, and Medium Resolution Spectroscopy.  The
plans for the three stages of MIRI data reduction are outlined.  The
1st stage processes the ramps into uncalibrated slopes, then the
slopes are calibrated in the 2nd stage, with the final stage focusing
on combining individual exposures and extracting source information.
The details of the data reduction are expected to evolve as the
analysis of ground and commissioning observations improves the
understanding of the MIRI detectors and instrument.  Finally, we have
given examples of the data and data reduction products expected for
each of the four MIRI Observation Templates.

\section{Acknowledgments}
The work presented is the effort of the entire MIRI team and the enthusiasm within the MIRI partnership is a significant factor in its success. MIRI draws on the scientific and technical expertise many organizations, as summarized in Papers I and II. 
A portion of this work was carried out at the Jet Propulsion Laboratory, California Institute of Technology, under a contract with the National Aeronautics and Space Administration.

We would like to thank the following National and International
Funding Agencies for their support of the MIRI development: NASA; ESA;
Belgian Science Policy Office; Centre Nationale D'Etudes Spatiales;
Danish National Space Centre; Deutsches Zentrum fur Luft-und Raumfahrt
(DLR); Enterprise Ireland; Ministerio De Economi{\'a} y Competividad;
Netherlands Research School for Astronomy (NOVA); Netherlands
Organisation for Scientific Research (NWO); Science and Technology Facilities
Council; Swiss Space Office; Swedish National Space Board; UK Space
Agency.

\clearpage

\begin{deluxetable}{ll}
\tablewidth{0pt}
\tablecaption{Imaging OT User Options \label{tab_imaging_template}}
\tablehead{\colhead{Question} & \colhead{Choices} }
\startdata
Source Coordinates & variety of coordinate systems \\
TA Filter\tablenotemark{\dagger} & F560W, F1000W, F1500W, or FND \\
TA Flux in Filter\tablenotemark{\dagger} & in $\mu$Jy \\
Subarray & SUB64, SUB128, SUB256, \\
  & BRIGHTSKY, or FULL \\
Dither Pattern & Cycling, Gaussian, Reuleaux, or None \\
Subpixel Sampling & Yes or No \\
Filter & F560W, F770W, F1000W, F1130W, \\
 & F1280W, F1500W, F1800W, F2100W, \\
 & and/or F2550W \\
Exposure Time & in seconds \\
\enddata
\tablenotetext{\dagger}{Only required for precision synoptic photometry using the SUB64 subarray.}
\end{deluxetable}

\clearpage

\begin{deluxetable}{ll}
\tablewidth{0pt}
\tablecaption{Coronagraphy OT User Options \label{tab_coronagraphy_template}}
\tablehead{\colhead{Question} & \colhead{Choices} }
\startdata
Source Coordinates & variety of coordinate systems \\
TA Filter & F560W, F1000W, F1500W, or FND \\
TA Flux in Filter & in $\mu$Jy \\
Coronagraph/Filter & F1065C, F1140C, F1550C, \\
 & and/or F2300C \\
Exposure Time & in seconds \\
Reference source & all information except choice \\
  & of coronagraphs  \\
\enddata
\end{deluxetable}

\clearpage

\begin{deluxetable}{ll}
\tablewidth{0pt}
\tablecaption{LRS Template User Options \label{tab_lrs_template}}
\tablehead{\colhead{Question} & \colhead{Choices} }
\startdata
Source Coordinates & variety of coordinate systems \\
TA Filter & F560W, F1000W, F1500W, or FND \\
TA Flux in Filter & in $\mu$Jy \\
Slit Type & Slit or Slitless \\
Dither Pattern & Point, Extended, or None \\
Exposure Time & in seconds \\
\enddata
\end{deluxetable}

\clearpage

\begin{deluxetable}{ll}
\tablewidth{0pt}
\tablecaption{MRS template options \label{tab_mrs_template}}
\tablehead{\colhead{Question} & \colhead{Choices} }
\startdata
Source Coordinates & RA \& DEC \\
TA Filter & F560W, F1000W, F1500W, or FND \\
TA Flux in Filter & in $\mu$Jy \\
Grating Setting & A, B or C \\
Dither Pattern & None, 4-point, 6-point, \\
  & or cycling pattern \\
Exposure Time & in seconds  \\
\enddata
\end{deluxetable}

\clearpage

\clearpage

\clearpage

\clearpage

\clearpage

\clearpage

\clearpage

\clearpage

\clearpage

\clearpage

\clearpage

\end{document}